\documentclass{arXiv-ijam}

\usepackage{graphicx}
\usepackage{empheq}

\def\thebibliography#1{
 \list
 {[\arabic{enumi}]}{\settowidth\labelwidth{[#1]}\leftmargin\labelwidth
 \advance\leftmargin\labelsep
 \usecounter{enumi}}
 \def\newblock{\hskip .11em plus .33em minus -.07em}
 \sloppy
 \sfcode`\.=1000\relax}

\makeatletter
\renewcommand\@biblabel[1]{}
\makeatother

\begin{document}

\markboth{}{}



\title{Non-equilibrium evaporation/condensation model}

\author{B.V. Librovitch}
\address{Naval Architecture Ocean and Marine Engineering,
University of Strathclyde, Henry Dyer Building,
100 Montrose Street, Glasgow G4 0LZ, UK}

\author{A.F. Nowakowski}
\address{SFMG, Department of Mechanical Engineering, The University of Sheffield,
\\
Mappin Street S1 3JD Sheffield, UK}

\author{F.C.G.A. Nicolleau\footnote{
Corresponding author.}} 
\address{SFMG, Department of Mechanical Engineering, The University of Sheffield,
\\
Mappin Street S1 3JD Sheffield, UK}

\author{T.M. Michelitsch} 
\address{Sorbonne Universit\'es,
Universit\'e Pierre et Marie Curie (Paris 6)
Institut Jean le Rond d'Alembert, CNRS UMR 7190,
4 place Jussieu
75252 Paris cedex 05,
France
}

\maketitle


\newcommand{\fig}{Fig.~\ref}
\newcommand{\sm}{\sum\limits}
\newcommand{\B}[1]{{\bf #1}}
\newcommand{\C}[1]{{\cal #1}}
\newcommand{\D}{{\tt d}}


\begin{abstract}
%
A new mathematical model for non-equilibrium
evaporation/condensation including boiling effect is proposed.
A simplified differential-algebraic system of equations is obtained.
A code to solve numerically this differential-algebraic system has
  been developed. It is designed to solve both systems of equations with and without the
  boiling effect. Numerical calculations of ammonia-water systems with
  various initial conditions, which correspond to evaporation and/or
  condensation of both components, have been performed.  
It is shown that, although the system evolves quickly towards a quasi equilibrium state, it is necessary to use a non-equilibrium evaporation model to calculate accurately the evaporation/condensation rates, and consequently all the other dependent variables.

\end{abstract}

%

\section{Introduction}

The problem of phase transition, and in particular evaporation/condensation, is
one of the most important problems of modern technology. 
There are numerous
applications of this process in industry, for example, in refrigeration and chemical industry.

It is very common to use an equilibrium evaporation model which assumes that the concentrations of species in the gas phase are always at saturated conditions
[\cite{MAKEYEV1981,ZVEREV1989}].  
This approach is not only conceptually questionable - indeed if the gas is at saturated condition there is no evaporation
or condensation - but sometimes it can lead to significant numerical errors, such as the obtention of
negative concentrations in complex computer simulations.  Therefore, for
industrial problems, for example the modelling of an absorption refrigeration 
cycle, it is paramount to develop and use non-equilibrium
evaporation/condensation model   
[\cite{Remorov2005,Chernyak1995,Chernyak1989},
\cite{Ivchenko1987,Ytrehus1996,Young1991},
\cite{Young1993,Wang2005},
\cite{Wang2006,Krishnaswamy2006}.]

In the work of \cite{Ivchenko1987} the moment method was applied to
solve the problem of evaporation/condensation of a spherical droplet
immersed in a vapour-gas mixture. Calculating the moments from the collision
integrals he proposed a new procedure which allows the use of collision integrals
in the Boltzmann form. He obtained an analytical formula for the time dependence of the
droplet radius for any Knudsen number.

The kinetic theory of droplet evaporation has been developed by \cite{Chernyak1995}. He studied the evaporation/condensation of a single
aerosol particle suspended in a non-equilibrium gas mixture. In his earlier
work together with Margilevsky [\cite{Chernyak1989}] he developed a linear
theory of mass and heat transfer for aerosol particle evaporation. 

The condensation and evaporation of a single liquid droplet for an arbitrary
Knudsen number have been studied analytically by
Young [\cite{Young1991,Young1993}] who developed a new system of equations
describing evaporation/condensation of a small liquid droplet. His theory is
valid for polyatomic gases and, apart from the evaporation/condensation
coefficients, contains no additional empirical constants. 

The continued development of diesel and rocket engines generated a significant
interest in the understanding of aerosol behaviour. A good comprehensive review
about equilibrium and non-equilibrium droplet evaporation models has been
written by Miller et. al. [\cite{Miller1998}]. These authors argued that the
Langmuir-Knudsen law [\cite{Knudsen1915,Knudsen1934}] should be used for general
gas-liquid f\/low calculations because it incorporates the realistic
non-equilibrium evaporation/condensation behaviour prevailing in many practical
situations while not requiring more computational effort than other equilibrium models.

Comprehensive theoretical research in the field of film condensation in
micro-channels have been performed by Wang and
Rose~[\cite{Wang2005,Wang2006}].  Their model is based on fundamental
principles and takes into account surface tension, vapour shear stress
and gravity. The effect of a channel geometry was investigated for
different types of cross-sections. They found that there is
a significant heat-transfer enhancement by surface tension towards to
the channel entrance. 

This paper is organised as follows. In \S\ref{Development_of_the_mathematical_model} the main assumptions are
introduced and the framework for the derivation of a novel approach is
discussed. Section \ref{Boiling_Evaporation_Model} is focused on the extension of
the model to the boiling situation.
The work is completed by the derivation of an analytical solution of a simplified system and the numerical solution of a differential-algebraic system
in \S\ref{Numerical_and_Analytical_Results}. The Conclusions are
presented in \S\ref{Conclusions}.
\section{Mathematical model}
\label{Development_of_the_mathematical_model}

The main assumptions for the model are:

\begin{itemize}
\item There is thermal equilibrium between the two phases. Namely, the gas temperature is equal to the liquid temperature, $T_{g}=T_{l}=T$. 
\item There is mechanical equilibrium between the two phases, that is the gas and liquid pressures are equal. The pressure gradient due to gravity can be neglected.
\item The gas phase is considered as an ideal gas.
\item The liquid phase is assumed to be incompressible.
\item The Stefan flux is neglected. 
\item The detailed bubble generation mechanism is neglected in the boiling model. 
\item The zero-dimensional approach is used. 
\end{itemize}

\subsection{Non-equilibrium evaporation/condensation model}

Let us consider a container which contains both liquid and gas phases.
For the sake of simplicity, let us assume that there is only one chemical species in the
volume. The generalisation to the multicomponent mixture is straightforward.

If the temperature of the liquid is less than the boiling temperature, $T<T_{b}$,
evaporation will only occur at the interphase surface. 
By contrast to evaporation, condensation can take place not only at the liquid-gas
interphase, but at all surfaces including the surfaces of the chamber.
Furthermore, it is assumed that phase transitions can only take place at the
liquid surface.

In the case of equilibrium, the molecular f\/lux which leaves the liquid surface is
balanced by the molecular f\/lux coming to the liquid from the gas phase.   
Evaporation occurs if the molecular f\/lux leaving the liquid surface is
greater than the f\/lux coming from the gas phase. Conversely, condensation takes
place when the molecular f\/lux from the liquid surface is less than the f\/lux coming
from the gas phase. 
Thus,
\begin{equation}\label{ev_rate}
S_{ev}=\dot N_{l \to g}- \dot N_{g \to l},
\end{equation}
here $S_{ev}$ is the total evaporation/condensation molar rate (e.g. mol~s$^{-1}$). It is positive for
evaporation, negative for condensation, and zero at saturated (equilibrium)
conditions. $\dot N_{l \to g}$ and 
$\dot N_{g \to l}$ are respectively the
molecular f\/luxes from liquid to gas and from gas to liquid. 
From the gas molecular theory [\cite{Schroeder1999}], it is well-known that
\begin{equation}\label{gas_flux_eq_1}
\dot N_{g \to l}= \frac{\xi_{g} U_{m}}{4} \Gamma A,
\end{equation}
here $A$ is the area of the interphase surface where evaporation/condensation
takes place, $U_{m}$ is the average molecular velocity and
$\xi_{g}$ is the molar volumic concentration of the component in the gas
phase.
$\Gamma$ is an
accommodation coefficient, which represents the fact that not all the gas molecules
which hit the liquid surface penetrate the liquid. In fact, a significant part
of them bounces back into the gas phase. It is therefore obvious, that $\Gamma$ must be
positive and not more than unity, $\Gamma \in [0:1]$. If $\Gamma=1$, all the gas
molecules which hit the liquid surface penetrate into the bulk of the liquid, and
if $\Gamma=0$, there is no phase transition at all.

Generally, $\Gamma$ can be a function of temperature, pressure and the
chemical composition of the liquid phase
[\cite{Smirnov1997,Remorov2005,Morita2003}].  In this paper, for the sake of 
simplicity, it is assumed that $\Gamma$ is constant. The value of $\Gamma$ can
be evaluated using the experimental data obtained from a dynamic evaporation
experiment.

Using gas molecular theory [\cite{Schroeder1999}] the average molecular velocity
can be expressed as
\begin{equation}\label{av_molecular_velocity}
U_{m}=
\left(
\frac{8 R_{u} T}{\pi W}
\right)^{0.5},
\end{equation}
here $R_{u}$ is the universal gas constant, $T$ is the temperature of the
system and $W$ is the molecular weight of a component.

The molecular flux which leaves the liquid surface is determined by the
internal state of the liquid, namely, its temperature, pressure, etc. It does not
depend on the concentration of the component in the gas phase. In some sense
the liquid ``does not know'' whether it is at equilibrium conditions or not.
Thus, using the fact that at equilibrium the two fluxes (in and out of the liquid)
are balanced, the following relation can be written
\begin{equation}\label{gas_flux_eq}
\dot N_{l \to g}= \dot N_{g \to l}^{eq}=
\frac{\xi_{g}^{eq} U_{m}}{4} \Gamma A,
\end{equation}
where $\dot N_{l \to g}$ is the molecular f\/lux from gas to liquid at
saturation, and $\xi_g^{eq}$ is the concentration of the component in the gas
phase at saturation.

Thus, substituting (\ref{gas_flux_eq_1}) and (\ref{gas_flux_eq}) into (\ref{ev_rate}) and using
(\ref{av_molecular_velocity}) for the  average molecular velocity, one gets:
\[
S_{ev}=\Gamma A
\left(\frac{R_{u}T}{2\pi W}\right)^{0.5}
\left(\xi_g^{eq}-\xi_g\right).
\]
This is the well-known Hertz-Knudsen's formula
[\cite{Knudsen1915,Knudsen1934,Smirnov1997}] - except that the original formula was written in terms of pressures.

In the case of a multicomponent mixture, the above formula can be generalised for each component $i$
as
\begin{equation}\label{evaporation_source}
S_{ev,i}=\Gamma_{i} A
\left(\frac{R_{u}T}{2\pi W_{i}}\right)^{0.5}
\left(\xi_{g_i}^{eq}-\xi_{g_i}\right),
\end{equation}

From (\ref{evaporation_source}), it is easy to find that 
there are three possible situations:
\begin{itemize}
\item[1)] the evaporation of the $i$-th component occurs if
 $\xi_{g_i} < \xi_{g_i}^{eq}$,
\item[2)] the condensation of the $i$-th component takes place if  
 $\xi_{g_i}>\xi_{g_i}^{eq}$,
\item[3)] the equilibrium (saturation) for the $i$-th component takes place if
 $\xi_{g_i}=\xi_{g_i}^{eq}$.
\end{itemize}
It is worth noticing that in order to derive
(\ref{evaporation_source}), the assumptions mentioned at the
beginning of this chapter have not been used. Therefore,
formula~(\ref{evaporation_source}) is quite general and can be used even if
some of the above assumptions are not satisfied.  
The main assumption in deriving (\ref{evaporation_source})
is that the molecules velocity distribution in the gas phase is Maxwellian so that
(\ref{av_molecular_velocity}) is satisfied.

\subsection{Species balance equation}

If the system is closed with a constant volume, the budget equations for $i$-th species are given by
\begin{itemize}
\item Gas phase
\begin{equation}\label{gas_concervation}
\frac{d(\alpha \xi_{g_i})}{d t}=\frac{S_{ev,i}}{V}.
\end{equation}
where $\alpha$ is the gas volume fraction,
$V$ the total volume of the chamber, and $S_{ev,i}$ the total evaporation rate of the $i$-th component. 
The constant volume assumption is made as it holds for
absorption refrigeration cycles and for refrigeration applications in general.
\item Liquid phase 
\begin{equation}\label{liq_concervation}
\frac{d((1-\alpha) \xi_{l_i})}{d t}=-\frac{S_{ev,i}}{V}
\end{equation}
\end{itemize} 

\subsection{Energy balance equation}

The energy conservation equation for a constant volume vessel in terms of temperature can be written as
(the detailed derivation is given in appendix~\ref{appendix_A})
\begin{equation}\label{enth_cons}
\left[\alpha \sm_{i=1}^{n} \xi_{g_i} \, c_{p_{gi}}+(1-\alpha) \sm_{i=1}^{n} \xi_{l_i} \, c_{p_{li}}\right]
\frac{d T}{d t}
-\alpha\frac{d P}{dt}
+\sm_{i=1}^{n}\Delta h_{i} \frac{S_{ev,i}}{V}=
\frac{\dot{Q}}{V}.
\end{equation}
where $c_{pgi}$ and $c_{pli}$ are the molar heat capacity at constant pressure for the $i$-species for respectively the gas and liquid phase.
$\Delta h_{i}$ is the molar latent enthalpy for the 
$i$-species.
$\dot{Q}$ is 
the rate of heat transfer from the surroundings to the system and is given by
\begin{equation}
\dot{Q} = \lambda A_w (T_w - T ),
\end{equation}
where $T$ is the temperature of the system, $T_w$ is the temperature of the wall, $\lambda$
is the heat transfer rate coefficient, and $A_w$ is the surface area for heat transfer to the
calorimeter.
\subsection{Equations of state} 

In order to simplify the model, it is assumed that the liquid phase is
incompressible and the vapour phase behaves as an ideal gas. In this case for
the vapour phase the following equation is satisfied
\begin{equation}\label{igeal_gas_equation}
P=R_{u}T\sm_{i=1}^{n} \xi_{g_i}.
\end{equation}
The volume of the liquid phase is given by
\begin{equation}\label{eq_state_liquid}
\sm_{i=1}^{n} \xi_{l_i} \bar V_{l_i}=1.
\end{equation}
where $\bar V_{l_i}$ is the partial molar volume of the $i$-th species, which
is assumed to be constant.

\subsection{Phase equilibria relation}

In order to complete the evaporation/condensation model,
(\ref{evaporation_source}), it is necessary to express the concentration of
the components at the saturation condition, $\xi_{gi}^{eq}$ as functions of the
temperature and composition of the liquid phase.
For condensable species, Raoult's law [\cite{Smith2005}] is used
\begin{equation}\label{raoults_law}
y_{i}P=p_{i}(T)x_{i},
\end{equation}
where $p_{i}(T)$, $x_i$ and $y_i$
are respectively the vapour pressure, mole fraction in the liquid phase and mole fraction in the gas phase
of the $i$-th species.
The
temperature dependence of the vapour pressure is given by Antoine's
equation [\cite{Smith2005}]
\begin{equation}\label{antoines_equation}
\ln \left ( p_{i}(T) \right ) =D_{i}-\frac{B_{i}}{T+T^{a}_{i}},
\end{equation}
where $D_{i}$, $B_{i}$, and $T^a_{i}$ are material dependent empirical
constants, that are well-tabulated.

Using, (\ref{raoults_law}) together with the ideal gas equation
(\ref{igeal_gas_equation}), and the definitions of mole fractions in gas and
liquid phase, $y_{i}$, $x_{i}$, it can be shown that
\begin{equation}\label{conven_sat_concetration}
\xi_{g_i}^{eq}=\frac{1}{R_{u}T}
\exp\left(D_{i}-\frac{B_{i}}{T+T^a_{i}}\right)
\frac{\xi_{l_i}^{eq}}{\sm_{i=1}^{n} \xi_{l_i}^{eq}}.
\end{equation}
Thus, the relation between the concentrations of a component in the gas and
liquid phases is obtained.

\section{Boiling evaporation model}
\label{Boiling_Evaporation_Model}

The previous evaporation model
(\ref{evaporation_source})-(\ref{eq_state_liquid}) and
(\ref{conven_sat_concetration}) was developed using the assumption that there
is no boiling. In case when boiling takes place additional consideration is
required.

By definition at boiling evaporation takes place not only at
the surface of the liquid but also in the bulk of the liquid. 
To model this
phenomenon, additional bulk evaporation source terms $S_{ev,i}^{b}$ must be
introduced.  These terms describe the evaporation rate in the bulk of liquid.
Thus, to take into account boiling in  (\ref{gas_concervation}),
(\ref{liq_concervation}) and (\ref{enth_cons}) the term $S_{ev,i}/V$ must be
substituted by $(S_{ev,i}+S_{ev,i}^{b})/V$.

There are now $n$ new variables, $S_{ev,i}^{b}$ $i=1...n$ that have been introduced in our
system. Therefore, it is necessary to add $n$ relations to close the system. 
During boiling, bubbles are generated within the liquid bulk. They contain
a mixture of saturated gases. Therefore, it is reasonable to suggest that
the evaporation rate of a component is proportional to the component concentration at saturation in the gas phase:
\begin{equation}
\left \{
\begin{array}{ll}
S_{ev,1}^{b} & = \xi^{eq}_{g_1}
\\
S_{ev,2}^{b} & = \xi^{eq}_{g_2}
\\
\dots        &  \dots
\\
S_{ev,n}^{b} & = \xi^{eq}_{g_n}
\end{array}\right .
\label{concentration_ratio}
\end{equation}
It is easy to see that the above relation consists of $n-1$ equations,
consequently, one more relation is needed to close the system.

From our point of view, it is consistent 
to suggest that the total bulk evaporation rate is
proportional to the difference between the saturated pressure $P_{eq}$ and current pressure $P$
in the system
\[
S_{ev,t}^{b}=\sm_{i=1}^{n} S_{ev,i}^{b} \sim (P_{eq}-P) \mathcal{H}[P_{eq}-P],
\]
here $S_{ev,t}^{b}$ is the total bulk evaporation rate.  
$\mathcal{H}$ is the Heavyside function with $\mathcal{H}[0]=0$. 
If boiling is in a
developed stage, bubbles are generated in the whole bulk of the liquid.
Therefore, it is natural to assume that the total bulk evaporation rate is
proportional to the volume of boiling liquid. Thus,
\begin{equation}\label{tot_bulk_evap_rate_1}
S_{ev,t}^{b}=\sm_{i=1}^{n} S_{ev,i}^{b}=\zeta(T,P,\xi_{l_i}) V_{l}(P_{eq}-P)
\mathcal{H}[P_{eq}-P], 
\end{equation}
here the correction factor $\zeta(T,P,\xi_{l_i})$ has been introduced, which, in
general, can be a function of the temperature, pressure, and liquid composition. 
For the sake of simplicity, it is assumed  that $\zeta$ is constant in our model.
Generally, the value of $\zeta$ or its dependence on other parameters can be
found from an experiment.

After substituting the ideal gas equation (\ref{igeal_gas_equation}) and
the expression for the liquid volume $V_{l}=(1-\alpha)V$ in formula~(\ref{tot_bulk_evap_rate_1}), it is found that
\begin{equation}\label{tot_bulk_evap_rate_2}
\sm_{i=1}^{n} S_{ev,i}^{b}=\zeta R_{u} T (1-\alpha)V
\left(\sm_{i=1}^{n}\xi_{g_i}^{eq}(T)-\sm_{i=1}^{n}\xi_{g_i}\right).
\end{equation}
Owing to the nature of boiling, the bulk evaporation rate of each component must
take a non-negative value.

For a complete boiling model a condition for boiling is needed.  It is
well-known that boiling takes place when the saturated pressure of the mixture
is larger than the pressure in the system, $P^{eq}>P$.
Using the equation for ideal gas~(\ref{igeal_gas_equation}), the boiling
condition can be written as
\begin{equation}\label{boling_cond_2}
\sm_{i=1}^{n} \xi_{g_i}^{eq} > \sm_{i=1}^{n} \xi_{g_i}.
\end{equation}

\section{Numerical results}
\label{Numerical_and_Analytical_Results}

\subsection{Numerical results for a constant wall temperature}

All numerical results in the following sections are related to the behaviour of
a two component ammonia/water system. This kind of binary system is chosen owing to its importance for the refrigeration industry. 
\\[2ex]
Using the following characteristic scales based on the vessel properties:
\begin{equation}
\left \{
\begin{array}{ll}
L_{*} &=V^{1/3}
\\[1.5ex]
T_{*}& =T_{in},
\\[1.5ex]
t_{*}& =\frac{R_{u}}{
A_{w} \lambda}, 
\\[1.5ex]
m_{*} & =\frac{R_{u}^{3} T_{in}}{(A_{w} \lambda)^{2} V^{2/3}}
\end{array}
\right .
\end{equation}
the system of equations~(\ref{gas_concervation}),
(\ref{liq_concervation}), (\ref{enth_cons}),
(\ref{igeal_gas_equation}), (\ref{eq_state_liquid}) and
(\ref{conven_sat_concetration}) can be written in the dimensionless form:

\begin{equation}
\label{non_dim_system_boiling}
\left\{
\begin{array}{l}
\frac{d (\alpha \tilde \xi_{g_i})}{d \tilde t}  =
\tilde S_{ev,i} + \tilde S_{ev,i}^{b} 
\\
\frac{d ((1-\alpha) \tilde \xi_{l_i})}{d \tilde t}=
-\tilde S_{ev,i} - \tilde S_{ev,i}^{b}
\\ 
\left[\alpha \sm_{i=1}^{n} \tilde \xi_{g_i} \tilde c_{p_{gi}}
+(1-\alpha)\sm_{i=1}^{n} \tilde \xi_{l_i} \tilde c_{p_{li}} 
\right]
\frac{d \tilde T}{d \tilde t} 
-\alpha\frac{d P}{dt} 
+\sm_{i=1}^{n} \Delta \tilde h_{i} (\tilde S_{ev,i} + \tilde S_{ev,i}^{b})=
(\tilde T_{w}-\tilde T) \\
\tilde P=\tilde T \sm_{i=1}^{n} \tilde \xi_{g_i} \\
\sm_{i=1}^{n} \tilde \xi_{l_i} \tilde V_{l_i}=1 \\
\tilde S_{ev,i}=\Gamma_{i} \tilde S 
\left(
\frac{\tilde T}{2 \pi \tilde W_{i}}
\right)^{0.5}
(\tilde \xi_{g_i}^{eq} - \tilde \xi_{g_i}) \\
\tilde \xi_{g_i}^{eq} = \frac{1}{R_{u} \tilde T}
\exp\left( D_{i}-\frac{\tilde B_{i}}{\tilde{T}+\tilde{T^a_{i}}}\right)
\frac{\tilde \xi_{l_i}}
{\sm_{j=1}^{n} \tilde \xi{l_j}} \\
\tilde S_{ev,1}^{b}/\tilde S_{ev,2}^{b}=
\tilde \xi_{g_1}^{eq}/\tilde \xi_{g_2}^{eq} \\
\sm_{i=1}^{n} \tilde S_{ev,i}^{b}= 
\tilde \zeta \tilde T (1-\alpha)
\left(\sm_{i=1}^{n} \tilde \xi_{g_i}^{eq}(\tilde T)
-\sm_{i=1}^{n} \tilde \xi_{g_i}\right)
\mathcal{H}\left[\sm_{i=1}^{n} \tilde \xi_{g_i}^{eq}(\tilde T)
-\sm_{i=1}^{n} \tilde \xi_{g_i}\right]
\end{array}
\right.
\end{equation}
This is a set of differential-algebraic equations (DAE). 
DAE are encountered in a number of scientific disciplines in particular in equilibrium chemistry.
The mathematical background for these equations and the different numerical methods used for solving them are presented and analysed in [\cite{Brenan-et-al-1995}].
Based on these methods we have developed an in-house code to solve the equation set (\ref{non_dim_system_boiling}).

\subsubsection{Initial conditions}

It is easy to see that in our systems of
equation~(\ref{non_dim_system_boiling}), there are
$2n+2$ first derivatives. Namely: $\frac{d\tilde \xi_{g_i}}{d \tilde t}$,
$\frac{d\tilde \xi_{l_i}}{d \tilde t}$, $\frac{d\alpha}{d\tilde t}$ and~$\frac{d
  \tilde T}{d \tilde t}$. Therefore, for a two component system ($n=2$), it
is necessary to provide our systems with 6 initial conditions.

Thus, to complete the problem the following initial conditions must be
specified 
$\tilde \xi_{g_1}(0)=\tilde \xi_{g_1}^{0}$, $\tilde \xi_{g_2}(0)=\tilde
\xi_{g_2}^{0}$, $\tilde \xi_{l_1}(0)=\tilde \xi_{l_1}^{0}$, $\tilde \xi_{l_2}(0)=\tilde
\xi_{l_2}^{0}$, $\alpha(0)=\alpha^{0}$, $\tilde T(0)=\tilde T^{0}$.
It is necessary to note that the equation of state for the liquid
(\ref{eq_state_liquid}) must be always satisfied. Therefore, only the
concentration of one component can be specified arbitrary in the range
$[0:1/\tilde V_{l_i}]$, the other must be calculated from the liquid state
equation (\ref{eq_state_liquid}). In all the cases considered in this paper
the initial concentration of ammonia in liquid phase is fixed, and
$\xi_{l_1}=3 \times 10^{4}$~mol~m$^{-3}$. From the above consideration it is
obvious that the initial concentration of water in the liquid phase is fixed, and
taken as $\xi_{l_2}=2.11 \times 10^{4}$~mol~m$^{-3}$. This value was calculated from (\ref{eq_state_liquid}) using the values for the specific volumes of both liquids given in Table~\ref{properties}.

The initial value for gas volume fraction, $\alpha^{0}$ is also fixed for all
calculations in this paper and $\alpha^{0}=0.5$. The reason for this is that
this value represents the volume of gas/liquid and does not have any
significant effect on the behaviour of the system. Unless it is very close to
the limiting values $0$ and~$1$, which correspond to one component system with
liquid or gas respectively. In this paper these two cases when the two phase system
becomes a one phase system corresponding to a complete evaporation or condensation are not
considered. The initial temperature is also fixed for all considered cases,
and $T^{0}=335$~K.

Thus, only the initial concentrations of both components in the
gas phase will be varied together with the wall temperature of the system.

Generally, the initial conditions for the concentrations in the gas phase can be
written as 
$\xi_{g_1}^{0}=a_{1} \, \xi_{g_1}^{eq}(T^{0})$, 
$\xi_{g_2}^{0}=a_{2} \, \xi_{g_2}^{eq}(T^{0})$,
where $a_{1}$ and $a_{2}$ are non-negative constants.  
\\[2ex]
For our numerical calculations the following five cases are considered.
\begin{enumerate}
\item\label{case_1} Evaporation of both components,
$\tilde \xi_{g_1}^{0} = 0.5 \, \tilde \xi_{g_1}^{eq}(\tilde T^{0})$,
$\tilde \xi_{g_2}^{0} = 0.5 \, \tilde \xi_{g_2}^{eq}(\tilde T^{0})$.
In this case, both initial concentrations are less than the equilibrium
concentrations, therefore, the boiling condition~(\ref{boling_cond_2}) is
satisfied. Thus, boiling takes place during the whole evaporation process.

\item\label{case_2} Evaporation of one component and condensation of the other
component (without boiling), 
$\tilde \xi_{g_1}^{0} = 1.6 \, \tilde \xi_{g_1}^{eq}(\tilde T^{0})$, 
$\tilde \xi_{g_2}^{0} = 0.4 \, \tilde \xi_{g_2}^{eq}(\tilde T^{0})$.
In this case one of the initial concentration is less than the equilibrium
concentration, and the initial concentration of the second component is more
than the equilibrium value.  Therefore, initially, one component evaporates
and the other condenses during the process.
It is easy to see that these initial conditions do not satisfy
inequality~(\ref{boling_cond_2}). Therefore, there is no boiling.

\item\label{case_3} Evaporation of one component and condensation of the other
component (with boiling), 
$\tilde \xi_{g_1}^{0} = 0.4 \, \tilde \xi_{g_1}^{eq}(\tilde T^{0})$, 
$\tilde \xi_{g_2}^{0} = 1.6 \, \tilde \xi_{g_2}^{eq}(\tilde T^{0})$.
Inequality~(\ref{boling_cond_2}) is satisfied, so there
is boiling.
Interestingly, in this case, the first component
(ammonia) evaporates in the bulk because of boiling and at the
surface. The second component (water) evaporates in the bulk but
condenses at the surface. Therefore, the total rate of evaporation of
the second component can be positive or negative.

\item\label{case_4} Condensation of both components, 
$\tilde \xi_{g_1}^{0} = 1.5 \, \tilde \xi_{g_1}^{eq}(\tilde T^{0})$, 
$\tilde \xi_{g_2}^{0} = 1.5 \, \tilde \xi_{g_2}^{eq}(\tilde T^{0})$.
In this case the concentrations of both components are higher than the
equilibrium concentrations. Therefore, both components condensate during the
whole process. 

\item\label{case_5} 
Both components are initially in
equilibrium:
$\tilde \xi_{g_1}^{0} = \tilde \xi_{g_1}^{eq}(\tilde T^{0})$,
$\tilde \xi_{g_2}^{0} = \tilde \xi_{g_2}^{eq}(\tilde T^{0})$.
These initial conditions do not satisfy
inequality~(\ref{boling_cond_2}), therefore there is no boiling.
\end{enumerate}

\subsubsection{Characteristic times}

In the considered problems there are two characteristic times, namely:
i) a characteristic time related to evaporation/condensation,
$\tau_{ev}$, ii) a characteristic time related to heating up/cooling,
  $\tau_{h}$. 
Typically the evaporation characteristic time is much smaller than the heating
evaporation time, $\tau_{ev} \ll \tau_{h}$.

After a relatively long period of time, $t \gg \tau_{ev}$ the initial
concentrations of the components are `forgotten' and the behaviour of all the 
considered cases is almost identical. The behaviour of the system is
determined mostly by the wall temperature or consequently, by the heat f\/lux
into the system.
By contrast, for a short period of time, $t \sim \tau_{ev}$, the initial
concentrations are very important but the inf\/luence of the wall temperature (heat
f\/lux) is negligible.

Accordingly, two sets of calculations are performed and discussed in this
paper: one set of calculations is done for a short time
$t_{f}=0.2$~s and a second set is done for a long time, $t_{f}=6 \times 10^{3}$~s.

It is reasonable to distinguish two types of equilibria, a)~concentration
equilibrium, and b)~thermal equilibrium. For the concentration equilibrium the
concentration of the component is equal, or almost equal, to the saturated
concentration at the current temperature. The external heat flux is not zero,
so the system can gain or lose internal energy. It is obvious that the system
can be in concentration equilibrium in one component while simultaneously an
other component can have a phase transition. 
By contrast,
when there is a thermal equilibrium, the external heat flux is zero but the concentration of one component
is not equal to the saturated value. Therefore, evaporation/condensation of the
component takes place.

If the system is in concentration equilibrium in all components and
simultaneously in thermal equilibrium, then the system is in a total
equilibrium, or just in an equilibrium. In this situation all processes are
stopped, and the system will remain in such a state for an indefinite time.

\subsection{Numerical results for cases with short period of calculation}

It is possible to show that for a short period, ($t \sim \tau_{ev}$) the influence of the wall temperature (external heat flux)
is not significant. Therefore for all the  calculations in this subsection,
one value of the wall temperature has been used, $T_{w}= 270$~K.
%
\begin{figure}[h]
\includegraphics[width=0.95\textwidth]{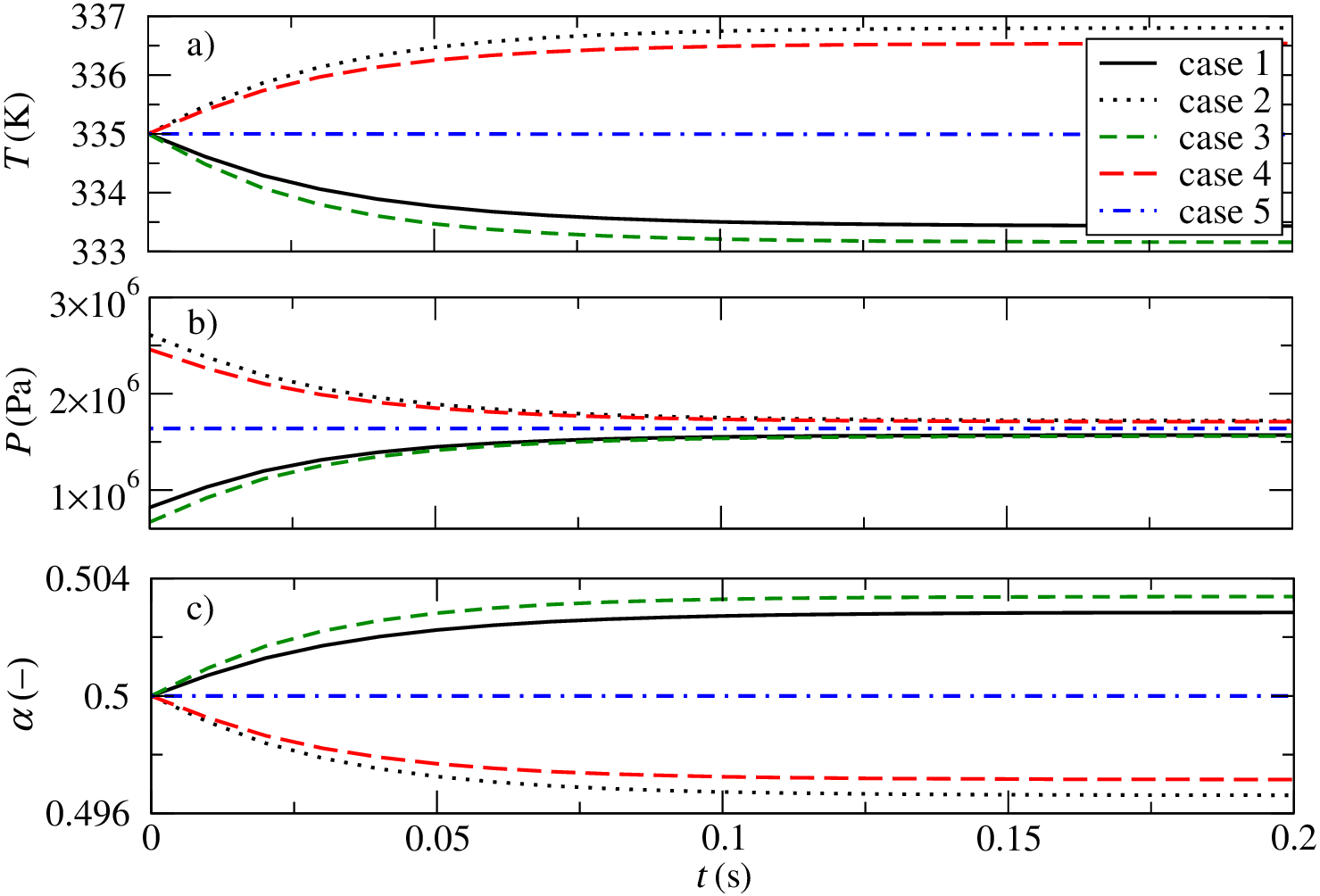}
\caption{\label{tem-pre-alpha} 
(a) Temperature, (b) pressure and (c) gas
volume fraction as functions of time for the short period of calculation, $t \sim \tau_{ev}$.}
\end{figure}
\\[2ex]
In \fig{tem-pre-alpha} profiles of temperature, pressure and gas volume
fraction are presented for all five cases under consideration.
It can be seen that all the profiles display a monotonic behaviour. 

In case~\ref{case_1} when the two components evaporate continuously with boiling,
temperature decreases because of evaporation, whereas pressure and gas volume
fraction increase. 
In case~\ref{case_2} ammonia (first component) condenses and water (second
component) evaporates without boiling, the temperature increases in time while 
pressure and gas volume ratio are reduced. In this case
there are two competing processes: the condensation of ammonia and the evaporation of
water.
In case~\ref{case_3} there is evaporation of ammonia and condensation of
water with boiling. It is worth emphasising that while there is boiling in
the system, simultaneously water condenses on the interface surface.
The temperature reduces in time whereas the pressure and gas volume
fraction increase.
In case~\ref{case_4} both components condense. The temperature increases in
time owing to the condensation of both components whereas the pressure and gas
volume fraction is reduced.
In case~\ref{case_5} the initial concentrations are at equilibrium and
therefore all the variables remain constant. Although, the external heat f\/lux is
not zero, as the wall temperature is different from the initial temperature of the
mixture, its inf\/luence is not significant for such short times.

\begin{figure}[h]
\includegraphics[width=0.95\textwidth]{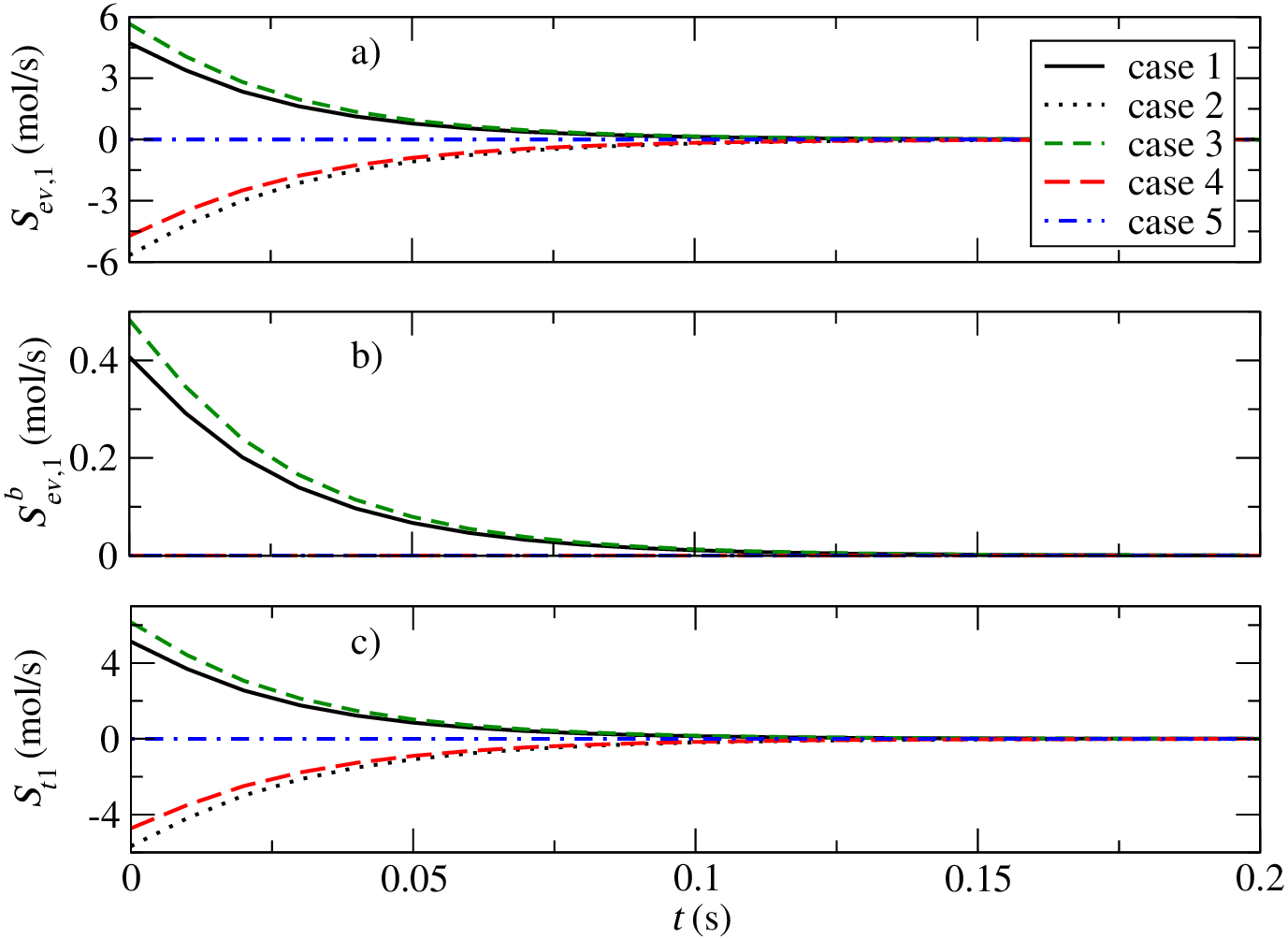}
\caption{\label{evap_rates_ammon} Ammonia evaporation rates: (a) at the
  surface, (b) in the bulk, and (c) the total evaporation rate as functions of
  time for the short time $t \sim \tau_{ev}$.}
\end{figure}

It is worth noting, that the pressures for all 5 cases do not approach the same
value. This is because for all 5 cases the initial conditions
for the gas concentrations are different, therefore, the total mass of the system
is different in each case. This causes the differences between  the concentration
equilibrium pressures.
%
\begin{figure}[h]
\includegraphics[width=0.95\textwidth]{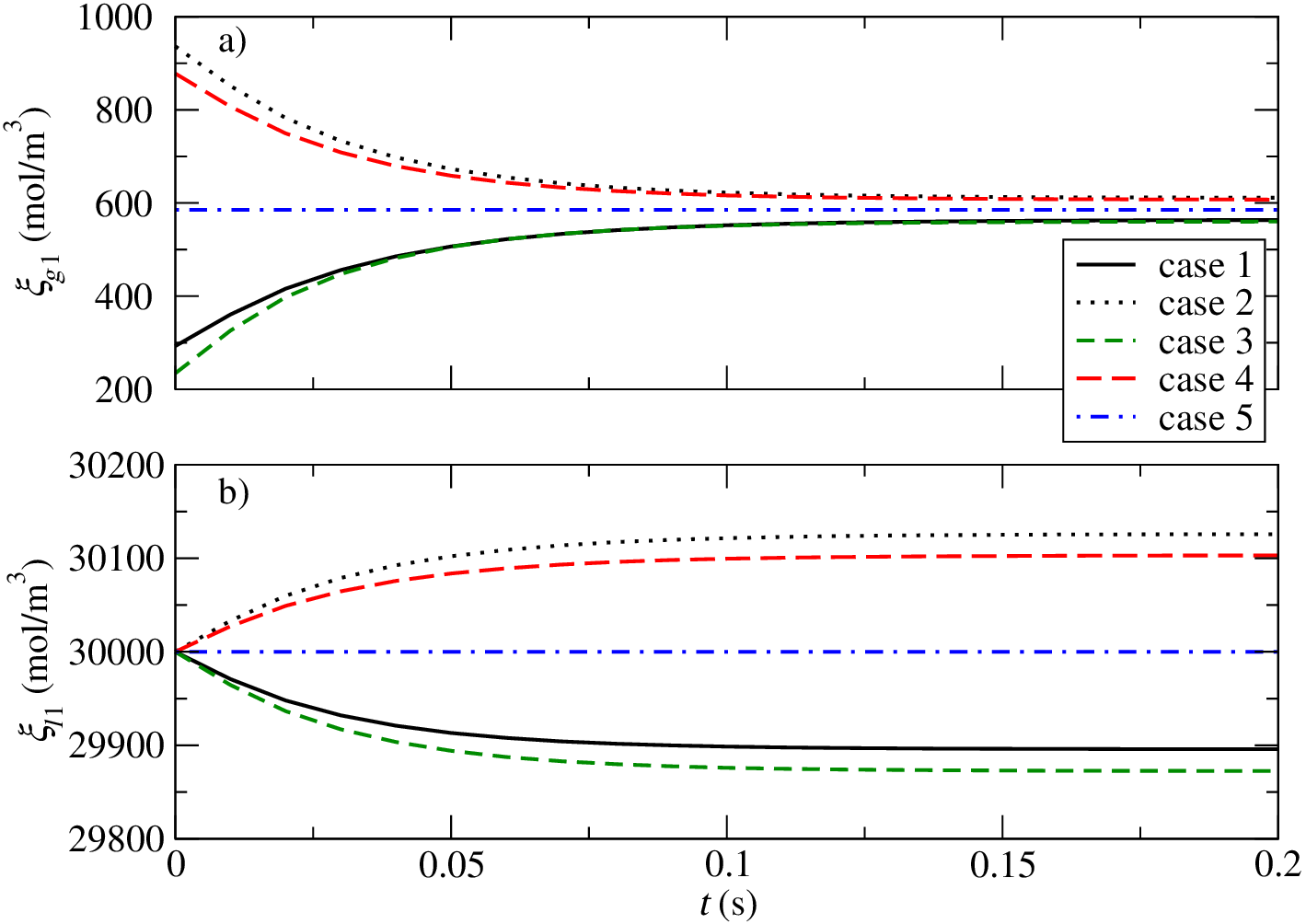}
\caption{\label{conc_ammon} (a) concentrations of ammonia in gas and (b)
  liquid phases as functions of time for the short time $t \sim \tau_{ev}$.}
\end{figure}
\\[2ex]
In \fig{conc_ammon} the gas (a) and liquid (b) concentrations of
ammonia are presented. 
The gas concentrations of ammonia and water do not approach a single value as the mixture temperatures are different in each case. 
\\[2ex]
All the evaporation rates (surface evaporation rate, bulk evaporation rate and
total evaporation rate) of ammonia are presented in
\fig{evap_rates_ammon}. 
%
As expected, all evaporation rate profiles monotonically approach zero,
which corresponds to the concentration equilibrium state.  The bulk evaporation
rate is positive for both components only for the two cases \ref{case_1} and
\ref{case_3} where boiling takes
place.

\subsection{Numerical results for cases with a large time calculation}

\begin{figure}[h]
\includegraphics[width=0.95\textwidth]{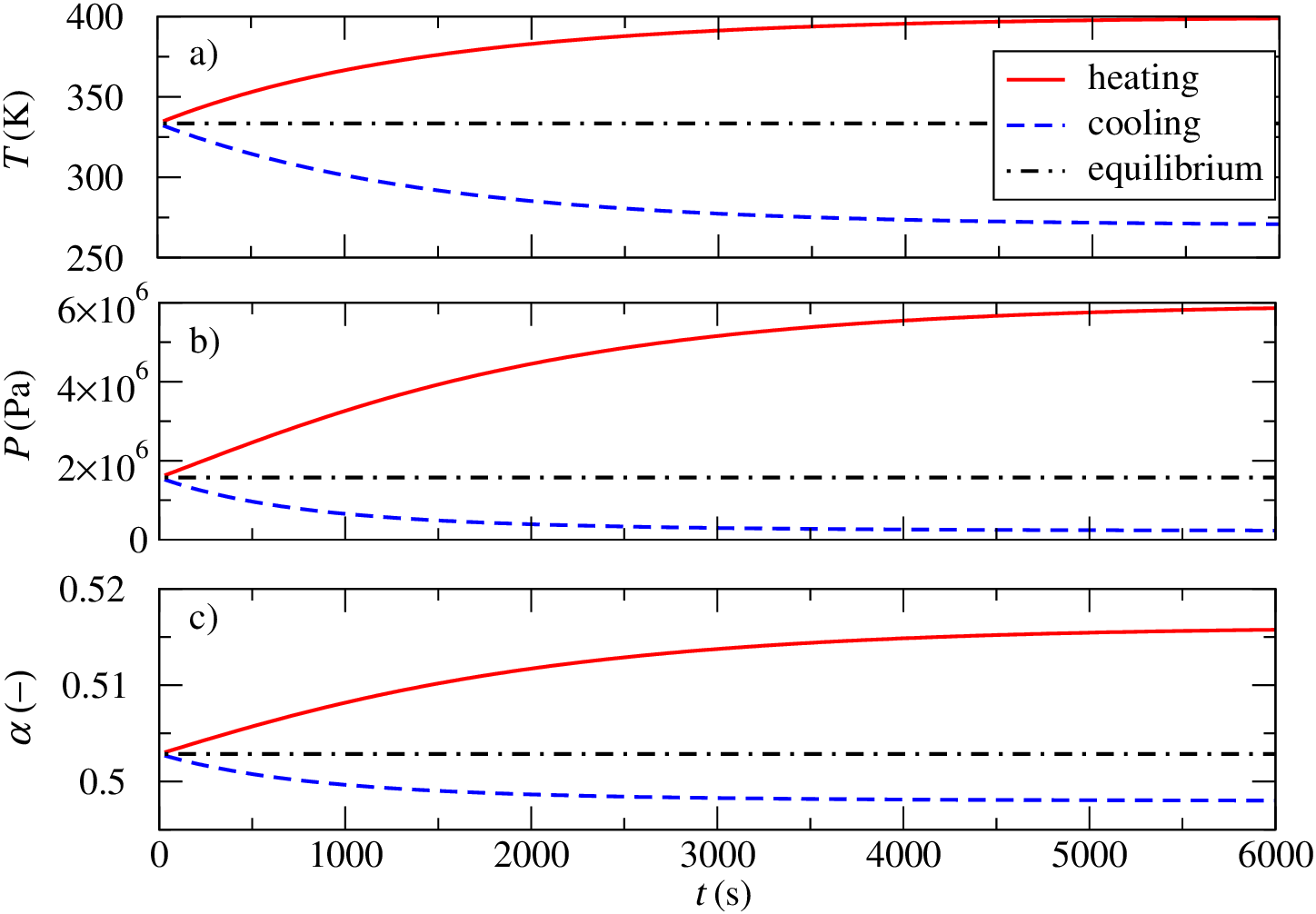}
\caption{\label{case_1_tem-pre-alpha} (a) temperature, (b) pressure and (c)
  gas volume fraction as functions of time for the long period
  of calculation $t \sim \tau_{h}$.}
\end{figure}

In the case of a large time of calculation ($t \sim \tau_{h}$) the initial
conditions for the gas concentrations are not so significant. For such long
time lags the external heat flux starts to play a significant role. Here, as
opposed to what happened in the previous subsection, the initial conditions for the gas concentrations are fixed and the wall temperature is varied in order to
investigate the system dependence on the external heat flux. Namely three cases
are considered: 
\begin{itemize}
\item[1)]heating, $T_{w}=400$~K. In this case the external heat
flux is positive, and the internal energy of the system is increasing in
time,  
\item[2)]cooling, $T_{w}=270$~K. The external flux is negative which causes
a decrease in the internal energy
\item[3)]
equilibrium, $T_{w}=335$~K. In this
thermal equilibrium the external heat flux is zero and the system is in an
equilibrium state.  
\end{itemize}
The initial conditions for the gas concentrations for all the cases in
this subsection correspond to the concentration equilibrium. They are fixed: $\tilde \xi_{g_1}^{0}=\tilde \xi_{g_1}^{eq}(\tilde T^{0})$, $\tilde
\xi_{g_2}^{0}=\tilde \xi_{g_2}^{eq}(\tilde T^{0})$.
In \fig{case_1_tem-pre-alpha} the temperature, pressure and gas
volume evolutions as functions of time are presented for the large time calculations.

\begin{figure}[h]
\includegraphics[width=0.95\textwidth]{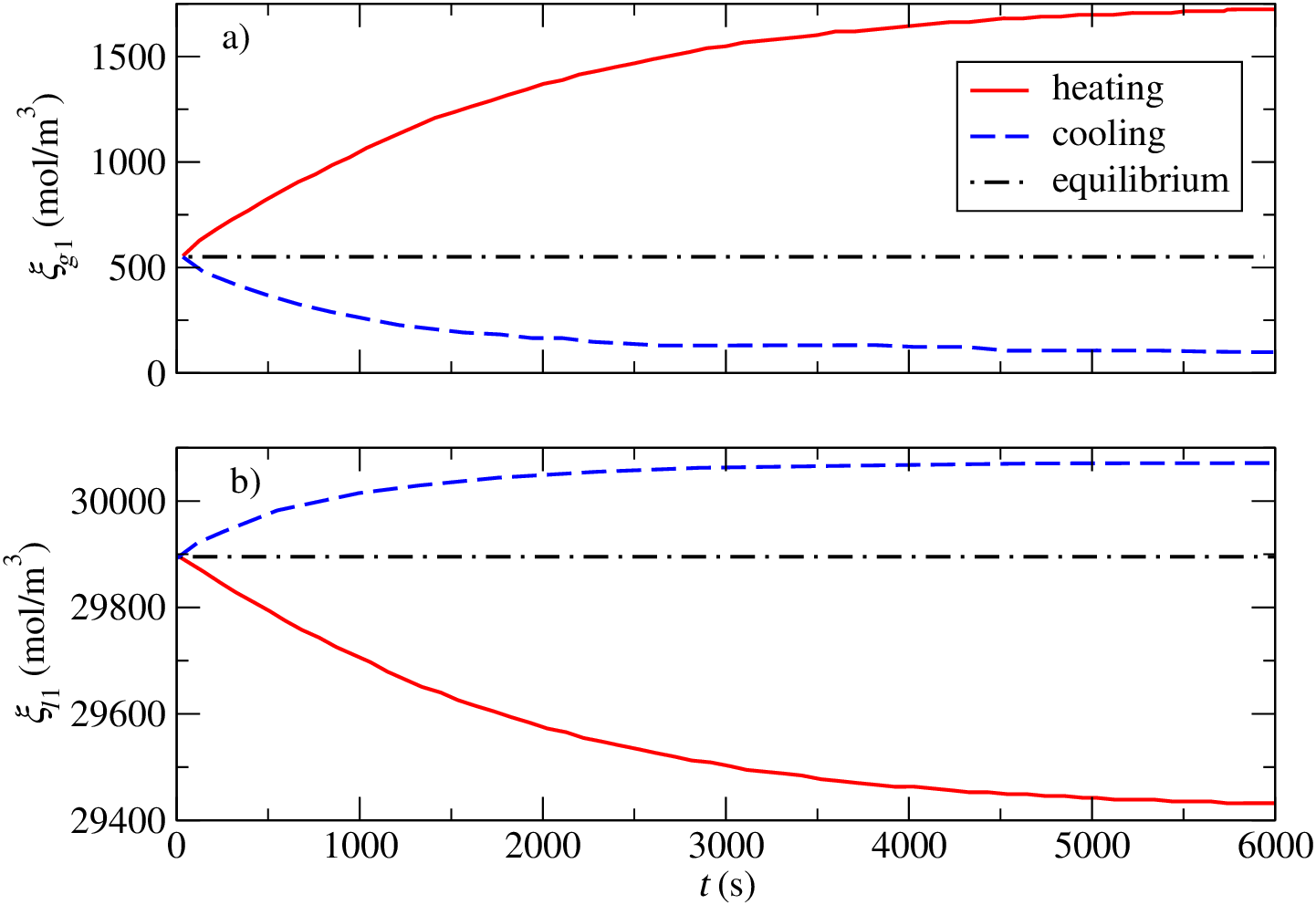}
\caption{\label{case_1_conc_ammon} Concentrations of ammonia in
  (a) gas and (b) liquid phases as functions of time for the long time
  calculation $t \sim \tau_{h}$.}
\end{figure}

The temperature of the system rises in time in the case of heating as the external
flux is positive. It approaches the wall temperature. The pressure also
rises as a consequence of heating, and the gas volume fraction increases as a consequence of evaporation.
In the case of cooling, the external heat f\/lux is negative. Therefore, in
contrast to the previous case, temperature, pressure and gas volume fraction
decrease during the process.
In the case of thermal equilibrium all the dependent variables remain constants and the  external heat flux is zero.
\\[2ex]
The concentrations of ammonia in both phases are
shown in \fig{case_1_conc_ammon}.
Owing to evaporation the gas concentration of ammonia is increasing and the 
liquid concentration is decreasing as would be expected.

In the case of cooling, the gas concentration of ammonia decreases and the
liquid concentration of ammonia increases in time. For water, both
concentrations in gas and liquid phases decrease.
\\[2ex]
\begin{figure}[h]
\includegraphics[width=0.95\textwidth]{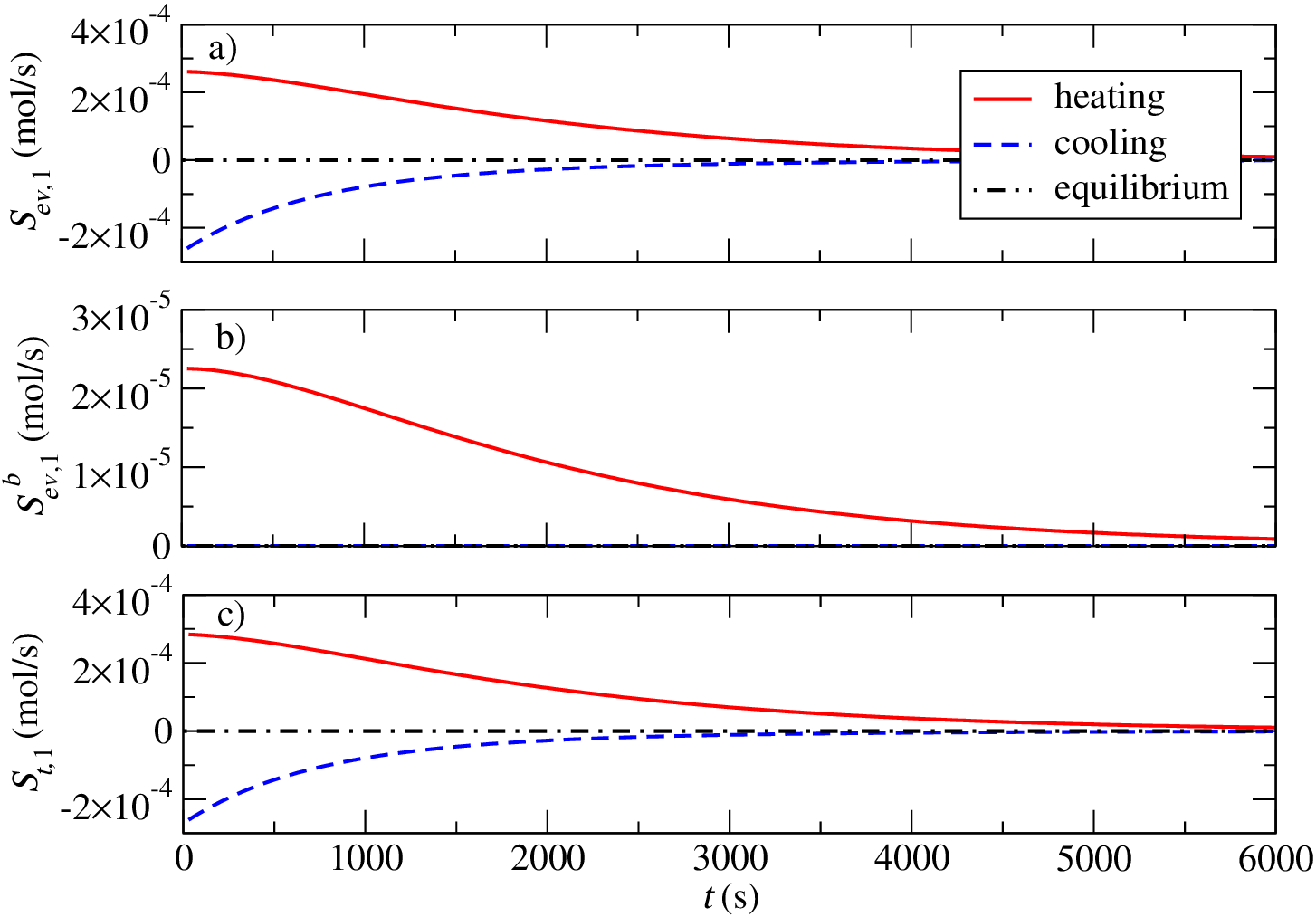}
\caption{\label{case_1_evap_rates_ammon} Ammonia evaporation rates: (a) at the
  surface, (b) in the bulk, and (c) the total evaporation rate as function of
  temperature for the long period of calculation, $t \sim \tau_{h}$.}
\end{figure}
All the evaporation rates for ammonia are
plotted in \fig{case_1_evap_rates_ammon}.  In the heating case there is boiling, so the bulk evaporation rate
is not zero. Whereas for the equilibrium and cooling cases the bulk evaporation rates
are zero.

It can be inferred from the plots that the ratio of the calculation periods for the long
and short cases is equal to $3 \times 10^{4}$. Thus, for the values of the
parameters we used, the system reaches a concentration quasi-equilibrium state after
a very short period of time (in our case it is approximately $0.1$~s). After
that it evolves relatively slowly to thermal equilibrium.

\section{Conclusions}
\label{Conclusions}

The main achievements of the work presented in this paper are:
\begin{itemize}
\item A novel mathematical model of non-equilibrium evaporation/condensation, including boiling, has been developed.
\item A new relationship (\ref{tot_bulk_evap_rate_1}) to close the system of
  equations with boiling has been proposed. It has been shown that this
  equation well describes the behaviour of the physical system. It only
  requires an additional parameter which can be obtained from an experiment.
\item A numerical code for the numerical solution of the differential-algebraic
  system has been developed. It was designed
  to solve both systems of equations with and without boiling and to switch
  from one regime to another, depending on the boiling
  condition~(\ref{boling_cond_2}).
\item Numerical calculations of an ammonia-water system with different initial
  conditions corresponding to evaporation and/or condensation of both
  components, and wall temperature have been performed.
\item It has been shown that, although the system quickly evolves to a quasi
  concentration equilibrium state (the differences between actual and
  equilibrium concentrations are rather small) it is necessary to use
  the non-equilibrium evaporation model, Eqs.~(\ref{evaporation_source}),
  (\ref{concentration_ratio}) and (\ref{tot_bulk_evap_rate_2}) to calculate
  the evaporation/condensation rates as well as all the other dependent variables
  accurately.
\end{itemize}


\appendix

\section{Derivation of the energy equation}
\label{appendix_A}

According to the 1st law of thermodynamics
the increase of the total internal energy of the system is equal to $Q$ the
heat input (or output) by the surroundings to the system or in terms of rates:
\begin{equation}\label{dU_dt}
\frac{d U}{dt}=\dot{Q}.
\end{equation}
It is assumed that the rate of heat transfer from the surroundings to the
system is given by
\[
\dot{Q}= \lambda A_{w}(T_{w}-T),
\]
where $T$ is the temperature of the system, $T_{w}$ the temperature
of the wall, $\lambda$ the heat rate transfer coefficient,
and $A_{w}$ the surface area for the heat transfer to the calorimeter.  The
total internal energy of the system can be written as
\[
U=H -P V=H_{g}+H_{l}-P V,
\]
where $H_{g}$ is the total enthalpy of the vapour phase and $H_{l}$ the total
enthalpy of the liquid phase. Substituting this into the energy balance
(\ref{dU_dt}), one can find if $V$ is constant:
\[
\frac{d H_{g}}{dt}+\frac{d H_{l}}{dt}
-V \frac{d P}{dt}=\dot{Q},
\]
Using the well-known thermodynamic relation, [\cite{Smith2005}]
\[
d H_{k}=C_{p_k}dT+V_{k}
\left[1
-\frac{T}{V_{k}}\left(\frac{\partial V_{k}}{\partial T}\right)_{P}
\right]dP+\sm_{i=1}^{n}H_{k_i} dN_{k_i},
\]
here the subscript $k=g,l$ stands for the gas or liquid phase,
the following relation is obtained
\begin{align*}
(C_{p_g}+C_{p_l})\frac{dT}{dt} 
+\left\{
V_{g}
\left[1-\frac{T}{V_{g}}
\left(\frac{\partial V_{g}}{\partial T}\right)_{P}\right]+
V_{l}
\left[1-\frac{T}{V_{l}}
\left(\frac{\partial V_{l}}{\partial T}\right)_{P}\right]
\right\}
\frac{dP}{dt} \\
+\sm_{i=1}^{n}H_{g_i} \frac{d N_{g_i}}{dt}
+\sm_{i=1}^{n}H_{l_i} \frac{d N_{l_i}}{dt}
-V\frac{dP}{dt}=\dot{Q},
\end{align*}
where $C_{p_g}$ is the overall heat capacity of the vapour phase,
$C_{p_l}$ is the overall heat capacity of the liquid phase,
$N_{g_i}$ is the number of moles of $i$-th species  in the vapour phase,
$N_{l_i}$ is the number of moles of $i$-th species  in the liquid phase,
$H_{g_i}$ is the partial molar enthalpy of the $i$-th species  in the
vapour phase, and $H_{l_i}$ the partial molar enthalpy of the
$i$-th species in the liquid phase.
Using the species balance equation leads to
\begin{align*}
(C_{p_g}+C_{p_l})\frac{dT}{dt} 
+\left\{
V_{g}
\left[1-\frac{T}{V_{g}}
\left(\frac{\partial V_{g}}{\partial T}\right)_{P}\right]+
V_{l}
\left[1-\frac{T}{V_{l}}
\left(\frac{\partial V_{l}}{\partial T}\right)_{P}\right]
\right\}
\frac{dP}{dt} \\
+\sm_{i=1}^{n}(\underbrace{H_{g_i}-H_{l_i}}
_{=\Delta H_{i}})
\underbrace{\frac{d N_{g_i}}{dt}}
_{=S_{ev,i}}
- V \frac{dP}{dt}=
\dot{Q}.
\end{align*}
For an ideal gas $\partial V_{g}/\partial T=V_{g}/T$,
so the first square brackets disappears.
For the liquid phase the isobaric thermal expansivity, $(\partial
V_{l}/\partial T)/V_{l}$, is small and can be neglected.  Therefore, the 
 equation can be considerably simplified
\[
(C_{p_g}+C_{p_l})\frac{dT}{dt}
-\alpha V \frac{d P}{dt}
+\sm_{i=1}^{n}\Delta H_{i} S_{ev,i}
=\dot{Q}.
\]
Substituting in the previous equation the following relations
\[
\begin{array}{cl}
C_{p_g}=\sm_{i=1}^{n}N_{g_i} \, c_{p_{g_i}}&=\alpha V \sm_{i=1}^{n} \xi_{g_i} \, c_{p_{gi}}\\
C_{p_l}=\sm_{i=1}^{n}N_{l_i} \, c_{p_{l_i}}&=(1-\alpha) V \sm_{i=1}^{n} \xi_{l_i} \, c_{p_{li}}
\end{array},
\]
where $c_{p_{gi}}$ and $c_{p_{li}}$ are the molar heat capacity of the $i$-th species in
gas and liquid phase respectively, then dividing by $V$, it becomes
\begin{equation}\label{enth_cons2}
\left[\alpha \sm_{i=1}^{n} \xi_{g_i} \, c_{p_{gi}}+(1-\alpha) \sm_{i=1}^{n} \xi_{l_i} \, c_{p_{li}}\right]
\frac{d T}{d t}
-\alpha\frac{d P}{dt}
+\sm_{i=1}^{n}\Delta H_{i} \frac{S_{ev,i}}{V}=
\frac{\dot{Q}}{V}.
\end{equation}

\section{Parameter estimation}
\label{appenB}

The values for the various physical properties that were used in the model
are summarised in Tables \ref{Parameters_Antoine_equation} and
\ref{properties} [\cite{Reid1987,Vargaftik1975,Forsythe2003,EngineeringToolBox}].

\begin{table}[h]
\tbl{Parameters for Antoine's equation (\ref{antoines_equation}).}
{\begin{tabular}{lccc}
\toprule
Species  & $D_{i}$ [--]& $B_{i}$ K & $T^a_{i}$ K\\
\colrule
Ammonia      & 22.40    & 2363.24    & -22.62     \\
Water        & 23.50    & 3992.51    & -38.48     \\
\botrule
\end{tabular}
\label{Parameters_Antoine_equation}}
\end{table}
%
\begin{table}[h]
\tbl{The physical properties of the species}
{\begin{tabular}{lcccc}
\toprule
Species & Molecular weight & $c_{p_{g,i}}$ & $c_{p_{l,i}}$ & $\bar V_{li}$ \\
  $i$    &  kg/mol  & J/(mol K)   &  J/(mol K)  & m$^3$/mol  \\
\colrule
Ammonia   & 0.017 & $37$  & $81$  & $2.065 \cdot 10^{-5}$   \\
Water     & 0.018 & $34$  & $75$  & $1.803 \cdot 10^{-5}$   \\
\botrule
\end{tabular}
\label{properties}}
\end{table}
The heat of vaporisation was estimated using the Clausius-Clapeyron
equation [\cite{Smith2005}]
\begin{equation}\label{heat_vaporisation}
\frac{\Delta h_{i}}{R_{u}} \approx
-\frac{d \ln(p_{i}(T))}{d (1/T)} =
\frac{B_{i}}{(1+T^a_{i}/T)^{2}}.
\end{equation}
All other parameters used to calculate
characteristic scales, are summarised here
$V=10^{-3}$~m$^{3}$, $A_{w}=6 \times 10^{-2}$~m$^{2}$,
$A=10^{-1}$~m$^{2}$, $\lambda=25$~J~m$^{-2}$s$^{-1}$K$^{-1}$. In all the
calculations the time step was fixed. For the short period calculations 
$\Delta t~=~10^{-3}$~s, while for the long period calculations $\Delta t= 10^{-2}$~s.

In the current model three empirical coefficients have been used. Namely: two
accommodation coefficients $\Gamma_{1}$ and $\Gamma_{2}$, and one correction
factor, $\zeta$ for total bulk evaporation rate at boiling
Eq.~(\ref{tot_bulk_evap_rate_2}).  For an accurate modelling, the values for
these coefficients should be estimated from experiments, which is beyond the
scope of this paper. For our calculations the following values
$\Gamma_{1}=\Gamma_{2}=10^{-1}$ and $\zeta=10^{-1}$~mol~s~m$^{-2}$kg$^{-1}$
have been used.




%
\bibliographystyle{apsrev}


%

\end{document}